\documentclass[12pt]{JHEP3}

\usepackage{cite}
\usepackage{enumerate}
\usepackage{amsbsy}
\usepackage[dvips]{graphicx}
\usepackage{graphics}

\newcommand{\be}{\begin{equation}} \newcommand{\ee}{\end{equation}}
\newcommand{\bea}{\begin{eqnarray}} \newcommand{\eea}{\end{eqnarray}}
\newcommand{\el}{\nonumber \\}
\newcommand{\re}[1]{(\ref{#1})}

\newcommand{\pat}{\partial}

\renewcommand{\sec}[1]{section \ref{#1}}

\newcommand{\brt}[1]{[#1]}
\newcommand{\para}{\paragraph}

\renewcommand{\a}{\alpha}
\renewcommand{\b}{\beta}

\newcommand{\GN}{G_{\mathrm{N}}}

\newcommand{\rmd}{\mathrm{d}}

\newcommand{\nonum}{\\}
\newcommand{\etal} {et al.}

\newcommand{\adot}{\pat_\tau{a}}
\newcommand{\addot}{\pat_\tau^2{a}}

\newcommand{\Hdot}{\dot{H}}

\newcommand{\Phidot}{\dot{\Phi}}
\newcommand{\Phiddot}{\ddot{\Phi}}
\newcommand{\Psidot}{\dot{\Psi}}
\newcommand{\Psiddot}{\ddot{\Psi}}
\newcommand{\bx}{\boldsymbol{x}}

\renewcommand{\H}{\frac{\pat_\tau a}{a}}
\newcommand{\HH}{\frac{(\pat_\tau a)^2}{a^2}}

\newcommand{\av}[1]{\langle{#1}\rangle}
\newcommand{\sQ}{\mathcal{Q}}
\newcommand{\sR}{{^{(3)}R}}

\newcommand{\PRD}[1]{{\it Phys. Rev.} {\bf D#1}}

\newcommand{\PRL}[1]{{\it Phys. Rev. Lett.} {\bf #1}}

\newcommand{\PLA}[1]{{\it Phys. Lett.} {\bf A#1}}

\newcommand{\MNRAS}[1]{{\it Mon. Not. Roy. Astron. Soc.} {\bf #1}}
\newcommand{\APJ}[1]{{\it Astrophys. J.} {\bf #1}}

\newcommand{\CQG}[1]{{\it Class. Quant. Grav.} {\bf #1}}
\newcommand{\GRG}[1]{{\it Gen. Rel. Grav.} {\bf #1}}
\newcommand{\AaA}[1]{{\it Astron. \& Astrophys.} {\bf #1}}
\newcommand{\PROG}[1]{{\it Prog. Theor. Phys.} {\bf #1}}

\newcommand{\IJMPA}[1]{{\it Int. J. Mod. Phys.} {\bf A#1}}
\newcommand{\IJMPD}[1]{{\it Int. J. Mod. Phys.} {\bf D#1}}

\title{Applicability of the linearly perturbed FRW metric and Newtonian cosmology}

\author{Syksy R\"{a}s\"{a}nen
\\ CERN, Physics Department Theory Unit \\ CH-1211 Gen\`eve 23, Switzerland \\
\\ Universit\'e de Gen\`eve, D\'epartement de Physique Th\'eorique
\\ 24 quai Ernest-Ansermet, CH-1211 Gen\`eve 4, Switzerland \\ \\
\email{syksy {\it dot} rasanen {\it at} iki {\it dot} fi}}

\abstract{It has been argued that the effect of
cosmological structure formation on the average
expansion rate is negligible, because the
linear approximation to the metric remains applicable
in the regime of non-linear density perturbations.
We discuss why the arguments based on the linear theory are
not valid. We emphasise the difference between Newtonian
gravity and the weak field, small velocity limit of general
relativity in the cosmological setting.}

\preprint{CERN-PH-TH/2010-048}

\begin{document}
  
\setcounter{tocdepth}{2}

\setcounter{secnumdepth}{3}

\section{Introduction} \label{sec:intro}

\para{The backreaction conjecture and perturbation theory.} 

It has been proposed that the observed increase in the expansion
rate and the distance scale of the universe at late times
relative to the matter-dominated homogeneous and isotropic
Friedmann-Robertson-Walker (FRW) model could be explained
by the breakdown of the homogeneous and isotropic approximation
because of the formation of non-linear structures
\cite{Buchert:2000, Tatekawa:2001, Wetterich:2001, Schwarz:2002, Rasanen}.
The effect of clumpiness on the average expansion rate is called
backreaction \cite{fitting, Buchert:1995, Buchert:1999}; see
\cite{Ellis:2005, Rasanen:2006b, Buchert:2007} for reviews.
The exact Buchert equations for the average expansion
rate show that large variance can lead to accelerated expansion as
faster regions come to dominate the volume \cite{Buchert:1999}.
This effect has been demonstrated with exact toy models
\cite{Rasanen:2006b, Chuang:2005, Paranjape:2006a, Kai:2006, Rasanen:2006a, Paranjape:2009}.
At late times there are deviations of order unity in the expansion
rate between different regions, so this mechanism could also
work in the real universe. The correct order of magnitude
and timescale of the change of the expansion rate have
been shown to emerge from the physics of structure formation in a
semi-realistic model without any free parameters \cite{Rasanen:2008a, peakrevs}.
The relation between the average expansion rate and observations
of light is also understood, though it should be established more rigorously and details remain to be worked out
\cite{Rasanen:2008b, Rasanen:2009a, Rasanen:2009b}.
However, there is no fully realistic calculation yet, and whether
backreaction is important in the real universe remains an open question.
The difference between Newtonian gravity and the weak field,
small velocity limit of general relativity
\cite{Ellis:1971, Ehlers:1991, Ellis:1994, Senovilla:1997, vanElst:1998, Ehlers:1998, Ehlers:1999, Szekeres}
plays an important part in the problem.
Therefore, quantifying the importance of the growth of structures
on the average expansion rate requires treating a statistically
homogeneous and isotropic but locally complicated non-linear
system in general relativity.

However, it has been argued that the effect of non-linear
structures on the expansion rate can be evaluated in
linear perturbation theory around the FRW metric
\cite{Seljak, Russ:1996, Wetterich:2001, Siegel:2005, Ishibashi:2005, Kasai:2006, Gruzinov:2006, Behrend, Paranjape, Paranjape:2009, Clarkson:2009a, Peebles},
sidestepping subtleties of non-linear general relativity
and the Newtonian limit.
The argument is that even though the density perturbation becomes
non-linear when structures form, the corresponding metric
perturbation in the longitudinal gauge, calculated from the
Poisson equation, remains much smaller than unity, so the effect
on the averages is negligible. There are multiple
problems with this argument. Evaluating the effect on averages
requires going at least to second order, so using first order
perturbation theory is inconsistent, observables are not given
by the metric alone, but by the metric and its derivatives
(which can become large) and, finally, the linear equations
do not, in fact, apply once the density perturbation becomes non-linear.
In short, it is not enough to calculate the magnitude of the effect
in linear perturbation theory, the applicability of the linear treatment
also has to be considered.

Some of these arguments have been addressed before
\cite{Buchert:1999, Rasanen, Kolb:2005a, Kolb:2005b, Rasanen:2006b, Rasanen:2008a, Buchert:2009, Rasanen:2009b}.
However, as they are being repeated in the literature,
it may be useful to discuss the issue in more detail than in
\cite{Rasanen, Rasanen:2006b, Rasanen:2008a, Rasanen:2009b},
and from a slightly different perspective than in
\cite{Buchert:1999, Kolb:2005a, Kolb:2005b, Buchert:2009}.
In section 2 we consider perturbation theory around the FRW metric
and show why the linear and second order calculations are not
sufficient for evaluating backreaction once the density field becomes
non-linear. We then look at the full non-linear equations
for the averages and consider the Newtonian limit.
In section 3 we discuss previous work on this topic,
and in section 4 we summarise our conclusions and outlook.

\section{Perturbations and the average expansion rate}

\subsection{The perturbative calculation} \label{sec:pert}

\para{The Einstein equation and the metric.}

We assume that matter and geometry are related by the
Einstein equation
\bea \label{Einstein}
  G_{\a\b} &=& T_{\a\b} \ ,
\eea

\noindent where $G_{\a\b}$ is the Einstein tensor
and $T_{\a\b}$ is the energy-momentum tensor; we
use units in which $8 \pi\GN=1$, where $\GN$ is Newton's constant.
We assume that the matter can be described as dust,
\bea \label{emt}
  T_{\a\b} = \rho u_\a u_\b \ ,
\eea

\noindent where $\rho$ is the energy density and $u^\a$ is the
velocity of the observers, taken to be comoving with the dust.

The perturbed FRW metric in the longitudinal gauge is
(we consider only a spatially flat background and only
scalar perturbations)
\bea \label{metric}
  \rmd s^2 = - ( 1 + 2 \Phi(t,\bx)) \rmd t^2 + ( 1 - 2 \Psi(t,\bx))\, a(t)^2 \delta_{ij} \rmd x^i \rmd x^j \ .
\eea

\noindent The Einstein tensor for the metric \re{metric} is
\bea \label{Ecomps}
  G^0_{\ 0} &\simeq& - 3 H^2 ( 1 - 2 \Phi ) - 2 a^{-2} \nabla^2 \Psi + 6 H \Psidot \el
  G^k_{\ k} &\simeq& - ( 2 \Hdot + 3 H^2 ) ( 1 - 2\Phi ) + 2 \Psiddot + 6 H \Psidot + 2 H \Phidot \el
  && - a^{-2} \nabla^2 (\Psi-\Phi) + a^{-2} \pat_k^2 (\Psi-\Phi) \el
  G^i_{\ j} &\simeq& a^{-2} \pat_i\pat_j ( \Psi -\Phi ) \qquad (i\neq j) \el
  G_{0i} &\simeq& 2 \pat_i ( \Psidot + H \Phi ) \ ,
\eea

\noindent where $\simeq$ denotes dropping terms which are higher
than first order (or, later, second order; it should
be clear from the context which is meant) in $\Phi$ or $\Psi$,
dot denotes derivative with respect to the background coordinate time $t$,
and no summation is implied in $G^k_{\ k}$.
We also split the velocity into the background and the perturbation,
$u^\a=\bar u^\a+\delta u^\a$, and assume that $\delta u^\a$
is small. From the normalisation condition $g_{\a\b} u^\a u^\b=-1$
it then follows that $u^0\simeq 1-\Phi$.

With the energy-momentum tensor \re{emt} and the Einstein tensor
\re{Ecomps}, the Einstein equation \re{Einstein} reduces, at first order, to
\bea
  \label{gen00} \!\!\!\!\!\!\! 3 H^2 ( 1 - 2 \Phi ) + 2 a^{-2} \nabla^2 \Psi - 6 H \Psidot &\simeq& \rho \\
  \label{genkk} \!\!\!\!\!\!\! 2 \Hdot + 3 H^2 - 2 \Psiddot - 6 H \Psidot - 2 H \Phidot + a^{-2} \nabla^2 (\Psi-\Phi) - a^{-2} \pat_k^2 (\Psi- \Phi) &=& 0 \\
  \label{genij} \!\!\!\!\!\!\! \pat_i\pat_j ( \Psi -\Phi ) &=& 0 \\ 
  \label{gen0i} \!\!\!\!\!\!\! 2 \pat_i ( \Psidot + H \Phi ) &\simeq& - \rho \delta u_i \ .
\eea

\noindent Note that we have not made any assumptions about the
perturbations of $\rho$.
From \re{genij} it follows that $\Psi-\Phi=A(t,x^1)+B(t,x^2)+C(t,x^3)$,
where $A,B$ and $C$ are arbitrary functions. We are mostly interested in
the situation when the perturbations are statistically homogeneous
and isotropic, in which case $A=B=C=0$, and we assume this
from now on. (The condition $\Psi-\Phi=0$ would
also follow from the technical requirement that the Fourier
transform of $\Psi-\Phi$ exists.)

\para{The static case.}

Let us first consider the static case $H=0$, and choose $a=1$.
The set of equations \re{gen00}--\re{gen0i} reduces to
\bea
  \label{static00} 2 \nabla^2 \Phi &\simeq& \rho \\
  \label{statickk} \Phiddot &=& 0 \\
  \label{static01} 2 \pat_i \Phidot &\simeq& -\rho\delta u_i \ .
\eea

To be consistent with neglecting terms which are second order
in $\Phi$, we should discard the right-hand side of \re{static01},
because according to \re{static00}, $\rho$ is of order $\Phi$.
We then obtain the result $\Phi=At+B(\bx)$, where $A$ is a
constant and $B(\bx)$ is determined by the density via \re{static00}.
It is possible for $\rho$ to have large variations without $\Phi$
becoming large or the first order treatment becoming invalid.
(This is the case in the solar system, for example).
However, in that case there is a slight inconsistency in the treatment,
because we have assumed that $\nabla^2\Phi=\frac{1}{2}\rho$ is small.
If $\rho$ is allowed to be large, we should equally treat
$\nabla^2\Phi$ as a large term, so products such as $\Phi\nabla^2\Phi$
should not be discarded. However, we should then take into account second
order terms in the metric, because they can be of the same order.
Let us look at this in more detail in the cosmological situation.

\para{The cosmological case.}

With $H\neq0$, the Einstein equation \re{gen00}--\re{gen0i} reads
\bea
  \label{cos00} 3 H^2 ( 1 - 2 \Phi ) + 2 a^{-2} \nabla^2 \Phi - 6 H \Phidot &\simeq& \rho \\
  \label{coskk} 2 \Hdot + 3 H^2 - 2 \Phiddot - 8 H \Phidot &\simeq& 0 \\
  \label{cos0i} 2 \pat_i ( \Phidot + H \Phi ) &\simeq& - \rho \delta u_i \ .
\eea

As is usual, we assume that the background and first order equations
are separately satisfied. This follows if we assume
that the average of $\Phi$ over the background space vanishes.
We split the density into the background value and the perturbation,
$\rho=\bar\rho+\delta\rho$, but do not assume that $\delta\rho$
is small. We then have
\bea
  \label{rhomean} 3 H^2 &=& \bar\rho \\
  \label{rhopert} 2 a^{-2} \nabla^2 \Phi - 6 H^2 \Phi - 6 H \Phidot &=& \delta\rho \\
  \label{Heq} 2 \Hdot + 3 H^2 &=& 0 \\
  \label{Phieq} \Phiddot + 4 H \Phidot &=& 0 \\
  \delta u_i &=& - \frac{2}{\rho} \pat_i ( \Phidot + H \Phi ) \ .
\eea

\noindent Equations \re{Heq} and \re{Phieq}, which come from
the pressure-free condition \re{coskk}, determine the evolution
of $a$ and $\Phi$, regardless of the energy density.
They lead to the standard relations $a\propto t^{2/3}$,
$\Phi=A(\bx)+B(\bx) t^{-5/3}$, where $A$ and $B$ are
arbitrary functions. According to \re{rhomean} and \re{rhopert},
the density contrast $\delta\equiv\delta\rho/\bar\rho$
is related to $\Phi$ by 
\bea \label{delta}
  \delta = \frac{2}{3 (a H)^2} \nabla^2 \Phi - 2 \Phi - \frac{2}{H} \Phidot \ .
\eea

We have nowhere required that $\delta$ should be small,
so one could at first sight think that \re{delta} applies
even when $\delta$ becomes of order unity, as long as
$\Phi$ remains small, analogously to the static case.
(In the static case $\bar\rho=0$, so $\delta$ is
not defined, but the variation of $\rho$ between
different regions of space can be large.)
Keeping to the linear theory, this is not true.
The time evolution of $\Phi$ is determined independently of
$\rho$, and inserting $\Phi=A+B t^{-5/3}$ into \re{delta}
shows that, dropping the decaying mode, the density contrast
has a constant part and a part which is proportional to $a$.
Therefore $\delta$ grows without limit. For an underdense
region, $\delta$ cannot go below $-1$, so this evolution is
clearly not correct as $\delta$ becomes
of order unity. The behaviour is also wrong for overdense
regions, as is well known from the spherical collapse model
\cite{Gunn:1972} (see \cite{spher} for reviews).

For the volume expansion rate $\theta=\nabla_\a u^\a$ we have
\bea \label{theta}
  \theta &\simeq& 3 H - 3 (\Phidot + H\Phi) + \pat_i u^i \el
  &=& 3 H \left( 1 - \frac{5}{3} \Phi - \frac{1}{3} \delta \right) \ ,
\eea

\noindent where we have on the second line dropped the decaying
mode of $\Phi$. (For the expression in terms of the proper time
measured by the observers, see \cite{Rasanen}.)
It is clear that the expansion rate given by the linear
theory is wrong when $\delta$ becomes of order unity.

\para{The reason for the breakdown.}

We have assumed that the Einstein tensor and the velocity $u^\a$ can be
expanded linearly in the metric perturbations. We have found
that the observables calculated using this procedure fail
to describe the real behaviour when $\delta\sim\nabla^2\Phi/(a H)^2$
becomes of order $\pm1$, even if $\Phi$ would seem to remain small,
so the linearly perturbed metric would appear to be valid.

The reason is that in neglecting all terms which are second
order in $\Phi$, we have implicitly assumed that terms such as
$\Phi\nabla^2\Phi/(a H)^2$ are much smaller than $\Phi$, i.e. that
$|\nabla^2\Phi/(a H)^2|\sim|\delta|\ll1$.
To extend the calculation into the regime $|\delta|\gtrsim1$,
we would have to expand to second order in $\Phi$. But to do
this consistently, we have to include the
intrinsic second order terms in addition to the squares of
first order terms. Indeed, the distinction between the two
is gauge-dependent \cite{Kolb:2004}, as 
first order quantities are not invariant under
second order gauge transformations. And at second order, the
metric cannot be written in the simple diagonal form \re{metric}
\cite{Matarrese:1997}.
It may be that the effect of the second order
terms is small for a particular quantity of interest,
but this has to be determined via a consistent calculation.

Effectively, there are three expansion parameters:
$\Phi$, $\pat_i\Phi/(a H)$ and $\nabla^2\Phi/(a H)^2\sim\delta$.
(No higher derivatives appear, because the Einstein equation is
second order.)
The formal perturbation expansion is defined in powers of
the metric perturbation, treated as an infinitesimal quantity
\cite{Bruni:1996}.
However, in the real universe, the metric perturbation
has a finite amplitude, so the gradients can make the other
expansion parameters large even when $\Phi$ remains small.
The gradient is a dimensional quantity, so a comparison scale
must enter. In cosmology, the relevant scale is
$a H$, and since $a H$ decreases in a decelerating FRW universe,
gradients become more important with time.
We can also view this as follows: for a time-independent $\Phi$,
the magnitude of $\nabla^2\Phi$ is fixed in time, while the
curvature scale of the universe, to which it is compared,
decreases. This kind of an instability is not present in the static case.

In a situation with multiple expansion parameters, perturbation
theory can be expected to remain valid when all
parameters are small, and to fail when all of them are large.
When some parameters become large while others remain small,
the validity of perturbation theory depends on the system,
and on the quantity under consideration.
In cosmology, the metric \re{metric} is simply the first order term
in an expansion, and when gradients of the metric perturbation become
large, higher order terms can no longer be neglected.
This does not necessarily mean that all first order results are wrong:
a consistent calculation with higher order terms may show that some
linear relations are valid. However, this cannot be determined
using the linear theory.

For backreaction, the important quantity is the average expansion
rate. (The primary quantities are of course observables defined
in terms of measurements of light; for the connection to the
average expansion rate, see \cite{Rasanen:2008b, Rasanen:2009a, Rasanen:2009b}.)
One might argue that even if the linear theory fails to correctly
describe the local quantities when density perturbations are
non-linear, the effect on the averages nevertheless remains small.
To address this issue, let us see what happens when we expand
to second order.

\para{The average expansion rate at second order.}

Taking the metric \re{metric} and calculating $\theta=\nabla_\a u^\a$
to second order in $\Phi$, we obtain (we take $\Phidot=0$; see
\cite{Rasanen} for the general expression)
\bea \label{locone}
  \!\!\!\!\!\!\!\!\!\!\!  \theta \simeq 3 H_{\tau} + \frac{118}{45}\frac{H}{(a H)^2}\pat_i\Phi\pat_i\Phi - \frac{2}{3} \frac{H}{(a H)^2} \pat_i \left( \pat_i\Phi + \Phi \pat_i\Phi - \frac{2}{3} \frac{1}{(a H)^2} \pat_i\Phi \nabla^2\Phi \right) \ ,
\eea

\noindent where $H_\tau\equiv2/(3\tau)$ is the background expansion
rate in terms of the proper time $\tau$ of comoving observers. 
The last term, with four derivatives, is of order $\delta^2$,
so there are large local variations in the expansion rate.
Averaging \re{locone} on the hypersurface of constant proper time,
we obtain \cite{Rasanen}
\bea \label{avone}
  \av{\theta} &\simeq& 3 H_{\tau} \left( 1 - \frac{22}{135}\frac{1}{(a H)^2}\av{\pat_i\Phi\pat_i\Phi}_0  + \frac{22}{27}\frac{1}{(a H)^2} \av{\pat_i (\Phi\pat_i\Phi)}_0 \right. \el
  && \left. + \frac{8}{27}\frac{1}{(a H)^4} \av{\pat_i \left(\nabla^2\Phi\pat_i\Phi\right)}_0 \right) \ ,
\eea

\noindent where $\av{}$ is a proper average with the correct
volume element, $\av{}_0$ is an average taken on the background
hypersurface of constant proper time, without perturbations in the
volume element, and we have assumed $\av{\Phi}_0=0$.
It is noteworthy that the term with four derivatives,
which has the largest amplitude locally, is a boundary
term. Before discussing this feature, let us note that
this calculation is not consistent, because we have
used the first order metric to calculate a second order
quantity, i.e. we have neglected intrinsic second order
terms. To obtain a result which does not depend on the gauge,
it is necessary to truncate the metric consistently
at second order instead of first order. The result is then
\cite{Kolb:2004}\footnote{As an aside,
in \cite{Kolb:2004} it is assumed, as is usual, that
the equations are satisfied separately order by order in
perturbation theory. While this procedure is self-consistent,
there seems to be no rigorous justification for it beyond first order.
At first order, the equations for the background and perturbations 
decouple, assuming that the average of the perturbations vanishes.
Starting at second order, the average of the
perturbations does not vanish, so decoupling of the
background and perturbations is an extra assumption.}
\bea \label{avtwo}
  \av{\theta} \simeq 3 H_\tau \left( 1 - \frac{5}{27}\frac{1}{(a H)^2}\av{\pat_i\Phi\pat_i\Phi}_0  + \frac{10}{27}\frac{1}{(a H)^2} \av{\pat_i (\Phi\pat_i\Phi)}_0 \right. \el
  \left. + \frac{16}{189}\frac{1}{(a H)^4} \av{\pat_i \left(\nabla^2\Phi\pat_i\Phi - \pat_i\pat_j\Phi \pat_j\Phi \right)}_0 \right) \ .
\eea

Comparing \re{avone} and \re{avtwo} shows that the first
order calculation in the longitudinal gauge happens to
give qualitatively the right answer, but the coefficients
of the terms are wrong.
(In first order perturbation theory, doing the
calculation in the synchronous comoving gauge, for example,
would give a qualitatively different result.)
Note that there is nothing in the result of the first order
calculation that would indicate that the answer is wrong.
An average of a total derivative can be converted into a
surface integral of a flux through the boundary.
If the distribution is statistically homogeneous and
isotropic, there is no preferred direction, so the
integral vanishes (up to statistical fluctuations).
(In perturbation theory, the technical requirement that
$\Phi$ can be expanded in Fourier modes would lead to the
same conclusion.)
With vanishing boundary terms, the correction to the mean
is $\sim\av{\pat_i\Phi\pat_i\Phi}_0/(a H)^2$, which is of
the order $10^{-5}$ for a realistic linear theory power spectrum.
To see the failure of the linear theory expanded to
second order, we have to work with the second order metric.
With the metric truncated at first order, it is impossible
to determine the magnitude of the higher order terms which
are neglected.
As the intrinsic second order terms are as large as the
first order terms squared, the question arises as to the
magnitude of the terms which are even higher order.
While calculating the coefficients of the various
terms would be an involved task, it is straightforward
to write down their general form.

\para{The general structure of the corrections.}

At second order, the possible correction terms
are the squares of the three expansion parameters, $\av{\Phi^2}_0$,
$\av{\pat_i\Phi\pat_i\Phi}_0/(a H)^2$ and
$\av{\nabla^2\Phi\nabla^2\Phi}_0/(a H)^4\sim\av{\delta^2}_0$.
(The quantity $\av{\Phi\nabla^2\Phi}_0$ is equal to 
$-\av{\pat_i\Phi\pat_i\Phi}_0$ up to a boundary term.)
For simplicity, we take $\Phidot=0$. As long
as $|\Phidot|\sim H|\Phi|$, taking into account
time-dependence would simply introduce more terms
of the same order of magnitude, and would not lead
to any qualitative change.
When the average expansion rate is expressed in terms
of the proper time $\tau$ (as opposed to the unphysical
coordinate time $t$), $\Phi$ appears in the expansion rate
only with derivatives acting on it \cite{Geshnizjani:2002}.
This is to be expected, because if $\Phi$ depends only
on time, it corresponds to using a different time
coordinate, not to having a physical degree of freedom.
Assuming that the higher order equations are satisfied order
by order and the perturbations are Gaussian, all higher
order terms from scalar perturbations factorise into
products of these three expectation values, since they
are sourced by the first order terms.
In contrast, vector and tensor perturbations (which
necessarily arise at higher orders) have solutions
which do not need to be supported by a source,
so their contribution cannot be completely expressed
in terms of the first order seed fields. (See
\cite{Matarrese:1997} for the second order case.)
However, such terms are expected to be subdominant to the
scalar perturbations in the non-linear regime.

We now return to the feature that at second order, the term
with the highest number of derivatives (and therefore locally
the largest amplitude) is a boundary term and as such vanishes
upon averaging, up to statistical fluctuations.
In \cite{Rasanen} it was argued that at higher
orders there might not be such a cancellation for the leading terms.
However, in \cite{Notari:2005} it was realised that because each
factor of $\pat_i/(a H)$ is accompanied by one power of the
speed of light $c$, the terms with the highest
number of spatial derivatives are the ones which dominate
in the Newtonian limit $c\rightarrow\infty$.
In Newtonian gravity, the backreaction correction is
exactly a boundary term \cite{Buchert:1995}.
Thus, in general relativity the term with the highest
number of spatial derivatives at each order in
perturbation theory is a boundary term\footnote{Assuming that
the leading order general relativity result reduces
to the Newtonian theory at all orders. As we discuss
in \sec{sec:Newton}, this is not necessarily true.
If that is not the case, the series \re{exp} is even more divergent.}.
The general structure of the corrections from scalar perturbations
to the average expansion rate is therefore (see also \cite{Kolb:2005a})
\bea \label{exp}
  \av{\theta} = 3 H_\tau \left( 1 + \frac{1}{(a H)^2} \av{\pat_i\Phi\pat_i\Phi}_0 \sum_{n=0}^{\infty} \lambda_n \av{\delta^2}_0^n + \ldots \right) \ ,
\eea

\noindent where $\lambda_n$ are constants and $\ldots$ indicates
subleading terms with a smaller number of derivatives,
such as $\av{\pat_i\Phi\pat_i\Phi}_0^m\av{\delta^2}_0^{n-m}/(a H)^{2m}$,
with $n\geq m\geq1$.
In powers of $\Phi$, the $\lambda_n$ term is of order $2n+2$.
The term $\av{\delta^2}_0$ can appear at fourth order in $\Phi$
at the earliest, where the leading correction is
$\av{\pat_i\Phi\pat_i\Phi}_0/(a H)^2\times\av{\delta^2}_0$.
This term grows without bound with increasing $|\delta|$, so
the breakdown of the perturbative expansion is transparent.
None of the coefficients $\lambda_n$ have been calculated.
It is possible to determine $\lambda_1$ in third order perturbation
theory, which is being developed \cite{Christopherson:2009},
without a full fourth order calculation \cite{Li:2007}.
However, calculating $\lambda_1$ would be inconclusive,
because at every order, there are an increasing number of
terms that grow even faster as $|\delta|$ becomes of order unity.

In \cite{Kolb:2005a} it was argued that the series \re{exp} would
have only a finite number of gradient terms when $\Phi$ is taken
as the full metric perturbation and not only the linear part
(and $\Psi\neq\Phi$ is included).
However, this is not the case: for a metric of the form \re{metric},
the velocity $u^i$ (and thus also $\theta$) expanded as a series
in $\Phi$ necessarily contains an infinite (or zero) number of
spatial derivatives \cite{Rasanen}.
And as we have noted, beyond first order, the metric cannot be
written in the form \re{metric} \cite{Matarrese:1997}.

From the fact that the series \re{exp} would naively seem to
diverge at $\av{\delta^2}_0\sim1$ we cannot conclude that the
sum of the correction terms would be large.
However, we can definitely say that the series expansion does
not prove that the correction would be small when $\Phi$ is small.
The magnitude of the effect has to be established with
non-perturbative methods, or a resummation of the series.
For studies in the spherically symmetric situation
where the exact solution is known, see
\cite{Paranjape:2009, Paranjape:2008, Kolb:2008, VanAcoleyen:2008, Enqvist:2009}.
In particular, \cite{Enqvist:2009} shows that it is
possible to have a large effect on the observable distance-redshift
relation even when the metric can be written in the form
\re{metric} (at least on the lightcone).
These models are not conclusive of the cosmological
situation, which is not spherically symmetric\footnote{Note that
in \cite{VanAcoleyen:2008} the average spatial curvature is small,
so it is clear that backreaction is not important.}.
Different resummation schemes have been applied in
Newtonian cosmology \cite{Carlson:2009}, and it would be 
interesting if such methods could be extended to
general relativity.

We would still expect to recover linear equations for
perturbations with wavelengths much larger than the
size of the structures, as is usual in statistical physics.
These should look similar to perturbation equations
around the FRW universe, with correction terms
due to the underlying structure \cite{Rasanen:2006b}.
This is also suggested by the success of FRW perturbation
theory in describing observations of large-scale structure.
Such equations would be analogous to the Buchert equations,
which look like FRW equations with correction terms, though
their physical content is different, as they involve only
average quantities and not local expansion.
The effect of backreaction cannot be described
merely as a change in the FRW background
\cite{Rasanen:2006b, Rasanen:2008a, Rasanen:2008b, Rasanen:2009b},
unlike argued in \cite{Kolb:2009, Clarkson:2009a, Clarkson:2009b}.
Even though the average expansion rate will always
agree with that of some FRW model, other observables
will in general not be the same as in that FRW universe.
In particular, the relationship between the average expansion
rate and the luminosity distance is different than in FRW models
if backreaction is important
\cite{Clarkson:2007b, Rasanen:2008b, Shafieloo:2009, Rasanen:2009b}.

We have discussed perturbation theory as it is most
commonly formulated, by adding perturbations on top of
a background (and previous perturbations).
The alternative is to take the full non-linear system
and linearise it.
In cosmology (unlike in the spherically symmetric case)
we cannot write down the exact solution to linearise.
However, it is at least possible to build perturbation theory
by starting from the full exact equations, written in the
covariant formalism, and linearise around
the FRW solution \cite{covpert, Ellis:1998c, Tsagas:2007}.
This has the benefit that all terms are included to begin with,
so it is transparent to estimate what is being dropped,
unlike in the case when perturbations are added to a background
order by order. In addition, the covariant
formalism deals only with measurable quantities and the
physical spacetime, so there are no gauge artifacts.
Instead of a perturbative analysis, we go directly
to the exact non-perturbative equations for physical insight
into the effect of perturbations becoming large.
A comparison of the general relativistic and Newtonian
cases is also instructive, given that backreaction vanishes
in the latter, for a statistically homogeneous and isotropic
distribution.

\subsection{The Newtonian limit} \label{sec:Newton}

\para{The Buchert equations.}

If the matter is irrotational dust, the exact equations
which describe the effect of inhomogeneities on the average
expansion rate $\av{\theta}\equiv 3\adot/a$ in general
relativity are \cite{Buchert:1999} (for the case with
non-dust matter or rotation, see
\cite{Buchert:2001, Larena:2009, Gasperini:2009b, Rasanen:2009b})
\bea
  \label{Ray} 3 \frac{\addot}{a} &=& - 4 \pi\GN \av{\rho} + \sQ \\
  \label{Ham} 3 \HH &=& 8 \pi \GN \av{\rho} - \frac{1}{2}\av{\sR} - \frac{1}{2}\sQ \\
  \label{cons} \pat_\tau \av{\rho} + 3 \H \av{\rho} &=& 0 \ ,
\eea

\noindent where $\sR$ is the spatial curvature and
$\sQ$ is the backreaction variable defined as
\bea \label{Q}
  \sQ \equiv \frac{2}{3}\left( \av{\theta^2} - \av{\theta}^2 \right) - 2 \av{\sigma^2} \ ,
\eea

\noindent where $\sigma^2$ is the shear scalar. The integrability
condition between \re{Ray} and \re{Ham} is
\bea \label{int}
  \pat_\tau {\av{\sR}} + 2 \H \av{\sR} = - \pat_\tau \sQ - 6 \H \sQ \ .
\eea

\noindent If $\sQ=0$, we have $\av{\sR}\propto a^{-2}$;
in particular, this holds for all exactly homogeneous and
isotropic universes \cite{Rasanen:2007}.
The system of equations \re{Ray}--\re{cons}
closes once we are given $\sQ$ or $\av{\sR}$; because of the
integrability condition \re{int}, the effect of
clumpiness can be viewed equivalently in terms of either quantity.
The Raychaudhuri equation \re{Ray} together with \re{Q} shows that,
apart from a possible $\av{\sR}\propto a^{-2}$ term,
deviations from homogeneity and isotropy have a large effect on
the average expansion rate only when the variance of the expansion
rate is large, and is not cancelled by the shear (or the shear
is large, and is not cancelled by the variance).
This shows that $\av{\delta^2}_0=1$ is not a sufficient
condition for a large effect on the average expansion rate.
However, it is necessary that the deviation of the expansion
rate from the mean is large in a large fraction of space
(assuming that the deviation is at most of the same order of
magnitude as the mean, which is true in cosmology).

\para{Newtonian gravity and beyond.}

In Newtonian gravity, the counterparts of the Raychaudhuri equation
\re{Ray} and the conservation equation \re{cons} are identical to
the relativistic equations, but there is no analogue of the
Hamiltonian constraint \re{Ham}. The variance of the
expansion rate and the shear combine to give a total derivative,
so $\sQ$ reduces to a boundary term \cite{Buchert:1995}.
(If the vorticity is non-zero, it is included in this
boundary term.) Thus, if backreaction is important in
a statistically homogeneous and isotropic universe,
this must be due to non-Newtonian aspects of general relativity
\cite{Notari:2005, Kolb:2005a, Kolb:2005b, Rasanen:2006b, Buchert:2007, Rasanen:2008a, Buchert:2008}.

In the expansion \re{exp}, all correction terms are
post-Newtonian. The term $c^2\av{\pat_i\Phi\pat_i\Phi}_0/(aH)^2$,
which may be identified as the square of a peculiar velocity,
$v^2/c^2$, suppresses the post-Newtonian terms.
However, the terms it multiplies can become very large as
$c^4\av{\nabla^2\Phi\nabla^2\Phi}_0/(a H)^4\sim\av{\delta^2}_0$ grows.
This demonstrates that in general relativity, non-Newtonian
effects can be important even when velocities are small
and fields are weak. An exact example is given by rotating
and expanding dust. In general relativity, there are no
dust solutions which have non-zero expansion and rotation
but zero shear \cite{Ellis:1967}.
However, in Newtonian gravity such solutions are exactly
known \cite{Ellis:1971, Senovilla:1997}.
Analysis of the Newtonian theory in this case would be
misleading, because the Newtonian solutions betray no sign of the
fact that starting from general relativity, they do not exist,
even at small velocities and weak fields\footnote{This
underlines that it is not sufficient to look at the Newtonian
limit of the equations of general relativity, but it is
necessary to consider the limit of solutions, because in general the
operations of taking the limit and solving the equations
do not commute \cite{Ehlers:1991, Senovilla:1997, Ehlers:1997}.}.

This issue arises due to the indeterminacy of Newtonian cosmology,
which is related to the absence of Newtonian analogues of the
magnetic component of the Weyl tensor\footnote{We can equivalently say
that in Newtonian gravity the magnetic component vanishes identically.}
and the evolution equation of the electric component of the Weyl tensor
\cite{Ellis:1971, Ellis:1994, vanElst:1998, Kofman:1995, Ehlers:1999, Szekeres, Ehlers:2009}.
Newtonian cosmology is only defined up to boundary conditions
\cite{Ellis:1994, Ehlers:1999, Szekeres}.
This shows up in the fact that $\sQ$ is a
boundary term, and the average expansion rate is determined by
what happens at the boundary. In general relativity this is not
the case, and backreaction is given by integrals over the volume.

If the volume considered has periodic boundary conditions or
is statistically homogeneous and isotropic, then in Newtonian
gravity $\sQ$ vanishes, and the first integral of the Raychaudhuri
equation \re{Ray} leads to an equation which looks like the
Hamiltonian constraint \re{Ham} with $\sQ=0$ and $\av{\sR}=E a^{-2}$,
where $E$ is a constant of motion which may be identified with
(being proportional to) the conserved energy of the isolated system.
In relativistic cosmology, the conserved energy is
replaced by spatial curvature, which has no physical analogue
in Newtonian gravity, so the interpretation of this term
is different in the two theories, even in the FRW case.
There is no conservation law for the spatial curvature in
general relativity, so $\av{\sR}$ can evolve in a non-trivial manner,
unlike the total energy of an isolated Newtonian system
\cite{Rasanen:2008a, Buchert:2008}.
The equivalent statement in terms of the backreaction variable $\sQ$
is that in Newtonian gravity the variance of the expansion rate
always equals three times average shear scalar (up to
a boundary term), as we see from \re{Q}, while in general
relativity there is no such constraint.

In second order perturbation theory we have, dropping
boundary terms and taking $\Phidot=0$,
\bea
  \av{\theta^2} - \av{\theta}^2 &\simeq& \frac{4}{9} \frac{H^2}{(a H)^4} \av{\nabla^2\Phi \nabla^2\Phi}_0 \approx \av{\delta^2}_0 H^2 \el
  \av{\sigma^2} &\simeq& \frac{4}{27} \frac{H^2}{(a H)^4} \av{\nabla^2\Phi \nabla^2\Phi}_0 \approx \frac{1}{3} \av{\delta^2}_0 H^2 \el
  \av{\sR} &\simeq& \frac{50}{9} \frac{1}{a^2} \av{\pat_i\Phi\pat_i\Phi}_0 \ ,
\eea

\noindent where the second approximation holds if $\Phi$ is small
compared to its gradient.
Second order relativistic perturbation theory around
a spatially flat FRW background is close to
Newtonian gravity in the sense that there is an exact
cancellation between the variance and the shear, so $\sQ=0$.
The variance and the shear can be calculated using first
order theory, because they vanish for the background \cite{Li:2007}.
There is only a single non-Newtonian term, the spatial curvature,
which is proportional to $a^{-2}$ according to the integrability
condition \re{int}. To determine the coefficient of this term,
it is necessary to go to second order.

Already in first and second order perturbation theory
the variance and the shear are large. A large backreaction
effect in general relativity does not require the variance
of the expansion rate to be larger than expected,
it is enough that the cancellation with the shear is not
perfect, unlike in Newtonian cosmology.
However, the spatial curvature does have to become large
\cite{Rasanen:2005, Rasanen:2006b, Rasanen:2008a}.
In a realistic cosmological setting, this is easy to understand.
The spatial curvature
of the initial overdense and underdense regions averages to zero
in the linear regime, but once perturbations become non-linear,
the evolution of overdense and underdense regions is different,
and the average will in general deviate from zero.
It is to be expected that if the volume of the universe
becomes dominated by underdense voids which expand faster than
overdense regions, the average spatial curvature will be negative.

Comparing Newtonian gravity and general relativity in
cosmology is different than in the case of isolated,
asymptotically flat systems. For isolated systems, both
Newtonian gravity and general relativity are well-defined.
In contrast, while relativistic cosmology is well-defined
there is no unique Newtonian theory of cosmology, because
the Newtonian equations are only defined up to
boundary terms which have to be specified at all times
\cite{Ellis:1994, Ehlers:1999, Szekeres}.
(We could alternatively say that Newtonian gravity is a theory of
isolated systems only, and there are an infinite number of
possible generalisations to the cosmological setting.)
This shortcoming of the Newtonian theory is often hidden
in cosmology by the assumption of periodic boundary conditions 
(sometimes implicitly through the use of Fourier series).
For periodic boundary conditions, Newtonian cosmology does have
a well-defined initial value problem \cite{Ehlers:1996, Ehlers:1999},
at least to all orders in perturbation theory, but this situation
does not correspond to the real universe, which is not periodic
on the observed scales.

So in the cosmological context, post-Newtonian corrections
refer to the difference between a realistic relativistic
cosmological model, which is thus far not tractable, and a
Newtonian model which is defined only when it does not
correspond to the real situation.
In particular, it is not possible to estimate backreaction
from usual N-body simulations, because the relativistic degrees
of freedom are absent. Relativistic cosmological
simulations would in principle provide an answer. The initial
conditions, the matter model and the equations of motion
are known and well-defined, but solving the system in full generality
is not computationally feasible. It would be interesting
to obtain a reduced system that would retain the relevant
relativistic degrees of freedom while being tractable.
Because relativistic cosmology is not sensitive to boundary
conditions in the same way as Newtonian cosmology, the
periodicity required for a numerical implementation would
not be a crucial limiting factor.

One check on the correctness of the Newtonian treatment
in the non-linear regime is provided by comparison of N-body
simulations with observations of structures. (In the simulations,
matter with negative pressure or modified gravity is introduced to change
the background expansion rate. Without such an addition, observations
of the expansion rate and the distance scale are, of course,
already completely discrepant with the Newtonian model.)
On large scales significant differences between the
simulations  and observations have been reported.
The observed homogeneity scale is an order of magnitude larger
than in N-body simulations \cite{SylosLabini},
and the number of very luminous superclusters is about five
times larger in observations than in simulations \cite{Einasto}.
Whether this reflects a deficiency of the Newtonian treatment,
or instead indicates a problem in the way simulations are done
or the observational data is analysed is not clear.
The discrepancy could also be due to an incorrect choice of
initial conditions, matter content or theory of gravity.

Relativistic dust models which are Newtonian-like are a
very restricted class \cite{silent}. The close relation of
linearly perturbed relativistic FRW models and Newtonian
gravity may be misleading because Newtonian-like models
suffer from a linearisation instability. 
In general relativity, the Newtonian constraint that the
magnetic component of the Weyl tensor vanishes identically 
is, in general, not propagated in time.
However, in the linear theory, the constraint is trivially
satisfied at all times. There are thus linear theory dust
solutions which are not the limit of any non-linear solution.
More importantly, this shows that relativistic dust models
do not, in general, have Newtonian counterparts and their
evolution cannot be described in Newtonian theory.
For practical applications in cosmology, the important
issue is the quantitative importance of the non-Newtonian
features, which depends on the solution under consideration.
For addressing this question it would be useful to
understand better the relation between the evolution of
the electric and magnetic components of the Weyl tensor
and spatial curvature in the context of cosmological
structure formation.

\section{Comparison to previous work}

\para{Arguments in linear theory.}

There have been various claims that backreaction in the
real universe is negligible
\cite{Seljak, Russ:1996, Wetterich:2001, Siegel:2005, Ishibashi:2005, Kasai:2006, Gruzinov:2006, Behrend, Paranjape, Paranjape:2009, Clarkson:2009a, Peebles}.
In particular, it has been argued that the relative magnitude of
backreaction corrections is given by the square of the peculiar velocity.
All these studies, except for \cite{Russ:1996, Clarkson:2009a},
expand quantities calculated with the first order metric to second order,
which, as we have seen, is not in general consistent.
Often the physical expansion rate, proper time and
hypersurface of averaging are also not correctly identified.
That was first done in \cite{Rasanen}, while the first
consistent second order calculation was done in \cite{Kolb:2004}.
Let us discuss some of the other shortcomings of these studies.

A numerical estimate of the correction to the expansion rate
from expanding the first order metric to second order was
first given in \cite{Seljak}.
In \cite{Russ:1996} the Zel'dovich approximation was
used to obtain the second order metric and calculate
corrections to the expansion rate.
In \cite{Wetterich:2001} the correction to the Einstein
equation was calculated in the same manner.
In \cite{Kasai:2006, Behrend, Clarkson:2009a} a calculation similar
to the one in \cite{Rasanen, Kolb:2004} was done, with some variations.
(In \cite{Behrend}, the perturbation of the volume element
was inconsistently neglected.)
In \cite{Siegel:2005} a similar calculation for the correction
to the 00-component of the Einstein equation was done, and
non-linear scaling relations were used for the density power
spectrum, but this cannot compensate for using only the linear metric.
As discussed in \sec{sec:pert}, such calculations lead to the
correction term $\av{\pat_i\Phi\pat_i\Phi}_0/(a H)^2$, which is
of the order $10^{-5}$. (In \cite{Russ:1996}, the value $10^{-3}$
was obtained instead.)
This result is the origin of the idea that the
magnitude of backreaction is given by the square of the
peculiar velocity, because at second order we have
$\av{\pat_i\Phi\pat_i\Phi}_0/(a H)^2\sim \av{u_i u^i}_0$.
However, beyond second order, the expansion parameter
$\av{\delta^2}_0$ is also involved,
so the second order calculation is inconclusive, and
$\av{\pat_i\Phi\pat_i\Phi}_0/(a H)^2\ll1$ is not a sufficient
condition for small backreaction.

As an aside, note that $u^i$ is the (spatial component of the)
deviation of the physical velocity of observers comoving
with the dust fluid from a fictitious background velocity.
This is a coordinate-dependent quantity, and we can always set
$u^i=0$ by choosing coordinates which are comoving with the observers.
In order to determine a physical peculiar velocity, we have to
define another physical velocity field to compare $u^\a$
to \cite{peculiar}. (In the longitudinal gauge in the linear
theory, $u_i u^i$ does give the physical magnitude of the
deviation from uniform motion.)

In \cite{Ishibashi:2005} it was asserted that the linear
metric \re{metric} (or the equivalent with a spatially curved
background) describes the universe on all scales, except in
the vicinity of black holes and neutron stars.
It was then claimed that if the conditions $|\Phi|\ll1,
|\Phidot|^2\ll a^{-2}\pat_i\Phi\pat_i\Phi,
(\pat_i\Phi\pat_i\Phi)^2\ll\pat_i\pat_j\Phi\pat_i\pat_j\Phi$
hold, non-linear corrections are negligible.
In fact, the metric \re{metric} cannot (with a dust source)
simultaneously describe the static metric in the solar
system and cosmological expansion, as is clear from the
expressions for $\theta$ and $\delta$ in \re{delta} and
\re{theta} together with \re{Phieq}. We can either
use the metric with $H=0$ to describe a static structure
or take $H\neq0$ to describe a cosmologically evolving
region, but these conditions are obviously mutually incompatible.
Apart from the inconsistency of using the linear theory
to calculate second order quantities,
if we nevertheless took \re{metric} and expanded
observables in terms of $\Phi$, then at fourth order and higher
we would expect to obtain terms involving $\av{\delta^2}_0$, which
are not necessarily small.
(Of course, these corrections are meaningless without accounting
for the intrinsic higher order terms.)

It was also argued in \cite{Ishibashi:2005} that since the average
expansion rate depends on the choice of the averaging hypersurface,
accelerated expansion could arise as a gauge artifact.
However, we should
distinguish three different concepts, namely gauge dependence,
coordinate dependence and dependence on the averaging hypersurface.
Gauge dependence arises due to ambiguity in the mapping between
the perturbed physical spacetime and a fictitious background spacetime.
When points of the fictitious and real spacetime
with the same coordinate values are taken to map to each other,
gauge dependence reduces to choice of coordinates, but in general
it is a distinct issue. In the covariant formalism with the
full non-linear equations, we deal only with physical quantities
and the real spacetime, so there is no gauge issue.
As all quantities are defined covariantly, independent of
the choice of coordinates, the dependence on the coordinate
system appears only in the usual transformation properties
of tensors under coordinate changes.
In particular, covariantly defined averages of scalar quantities
such as the volume expansion rate do not depend on the coordinate
system when expressed in terms of a physical observable such
as the observer's proper time \cite{Kolb:2004}.
They do, however, depend on the choice on the averaging hypersurface
\cite{Geshnizjani:2002, Geshnizjani:2003, Rasanen:2004}.
The reason is that the averaging hypersurface is physical
issue, unlike coordinates or gauge.
For irrotational dust, there is a preferred foliation
which is orthogonal to the fluid flow, and which coincides with
the hypersurface of constant proper time
\cite{Kolb:2005b, Rasanen:2006b, Rasanen:2008a}.
However, the hypersurface should be chosen based
on analysis of observables, and cannot be determined on abstract
mathematical grounds \cite{Rasanen:2006b, Rasanen:2008b, Rasanen:2009b}.
Any average quantities are of course useful only insofar they
give an approximate description of what is actually measured.
(For discussion of gauge-invariance in averaging, see also
\cite{Gasperini:2009a, Gasperini:2009b}.)

In \cite{Gruzinov:2006} the linear metric was again expanded
to second order. It was assumed that the average energy
density is the same as the background energy density, which is not
true beyond first order. Accordingly, one obtains equations
which are inconsistent \cite{Rasanen:2008a}. It was also argued
that backreaction vanishes in a 2+1-dimensional model.
This is not surprising, because in 2+1 dimensions, the 
integral of the spatial curvature is a topological invariant.
Therefore it cannot evolve in time, similarly to the total energy
of a Newtonian universe discussed in \sec{sec:Newton} \cite{Rasanen:2008a}.
Therefore, in 2+1 dimensions, backreaction can only give
$\sQ\propto a^{-4}$, and it is not clear whether the coefficient
can be non-zero. (This follows from the 2+1-dimensional analogue
of \re{int}: in $d+1$ dimensions, the last term
on the right-hand side is $-2d\H\sQ$.)

In \cite{Paranjape, Paranjape:2009} an iterative calculation was
done in the macroscopic gravity formalism \cite{Zalaletdinov},
which is an extension of general relativity. We are interested
in what happens in general relativity, but let us note that
the method of \cite{Paranjape, Paranjape:2009} was to take the first order
metric, expand to second order to obtain a new background
and then repeat the process. This way, one never moves beyond
first order perturbation theory. As we have seen, in general
relativity the higher order terms in general do not
have the same form as the first and second order terms, so this
kind of an analysis would not be correct there.

In \cite{Peebles} quantities were expanded to second
order in the linear metric, with the usual result.
The correction
$\av{\pat_i\Phi\pat_i\Phi}_0/(a H)^2\times\bar\rho\sim v^2\bar\rho$
was identified with a pressure term.
For clarity we note that the pressure measured by observers
comoving with dust is zero by definition.
The physical interpretation of the second order correction
is spatial curvature, not pressure.
To determine the pressure (and anisotropic stress
and energy flux) generated in the process of structure
formation, it is necessary to go beyond the ideal fluid treatment
\cite{McDonald:2009}. For the importance of non-dust terms
for backreaction, see \cite{Rasanen:2009b}.

Let us also comment on some studies which claim not a small,
but instead a possibly large backreaction effect from perturbation
theory.

A series expansion similar to \re{exp} was presented in
\cite{Notari:2005}. It was argued that the expansion
parameter is not $\av{\delta^2}_0$, but a quantity which becomes
of order unity around the present time. However, the expansion
in \cite{Notari:2005} is incorrect, because it does not take
into account the factorisation of higher order terms
into two-point functions and the constraint
that a non-zero two-point function must contain an even number
of momenta \cite{Rasanen:2008a}.
The only preferred era in the perturbative expansion is
$\av{\delta^2}_0=1$, signifying the formation of the first
generation of gravitationally bound objects.
For typical models of supersymmetric dark matter this
happens around a redshift of 40--60 \cite{SUSYCDM},
considerably earlier than the present day.
(As discussed above, the failure of the
simple perturbative expansion \re{exp} at $\av{\delta^2}_0=1$
does not alone indicate that backreaction would be large.)

In \cite{Li} first order theory expanded to second order was
used to estimate backreaction and compared to observations.
Apart from the question of the applicability of first
(or second) order perturbation theory, the correction
term used is qualitatively wrong, because the momentum
scale in the integral is misidentified with the size of the
averaging domain (the suppression of the leading correction
due to the fact that it is a boundary term is also neglected);
see section 5.1 of \cite{Rasanen:2008a}.

\section{Conclusion}

\para{Summary and discussion.}

If we consider cosmological perturbations which are initially
small and Gaussian with zero mean, it is necessary to go
at least to second order to find their effect on the average
expansion of the universe, called backreaction.
Nevertheless, taking the first order metric in the longitudinal gauge
and expanding quantities to second order serendipitously
gives almost the correct second order result \cite{Rasanen, Kolb:2004}.
At second order, the perturbations only lead to a spatial
curvature term with an amplitude of $10^{-5}$.
Several papers have thus claimed that backreaction is negligible
in the real universe, based on the argument that the linear
metric is a good approximation even when density perturbations
are non-linear.

We have discussed why the linearly perturbed FRW metric does
not in general correctly describe the situation once density
perturbations become non-linear.
The effect on the average expansion rate vanishes at the linear
level by construction, and at second order, the intrinsic second
order terms are of the same order as squares of the first order
terms (in fact, the division between the two is gauge-dependent).
At higher orders, generic correction terms become larger than
unity as density perturbations become non-linear.
This does not necessarily mean that the effect
on the average expansion rate is large, simply that the
naive perturbative expansion is no longer valid.

The important question is not in which form the metric can
be written, but what happens to measurable quantities.
For this purpose it is useful to consider the covariant
formalism, which deals only with physical
degrees of freedom and is fully non-linear.
The effect of deviations from homogeneity and isotropy
is quantified by the Buchert equations, which
show that the average expansion rate will significantly
deviate from the FRW behaviour when the variance of the
expansion rate is of order unity and does not cancel
against the shear (or vice versa) \cite{Buchert:1999}.
Even in the linear (and second order) theory, the variance
of the expansion rate becomes of order unity as
density perturbations become non-linear.
However, there is no significant backreaction
in relativistic second order theory because the variance
cancels exactly against the shear apart from a boundary
term, a feature shared by non-linear Newtonian gravity.
In exact general relativity, there is no such cancellation,
so Newtonian theory is not sufficient for evaluating backreaction.

Determining whether structure formation in the real universe
leads to a large enough variance for backreaction to be
important requires dealing with a locally
complex non-perturbative system in general relativity.
However, details of the local behaviour are not needed, statistical
information about the distribution of the expansion rate in
different regions is enough.
A semi-realistic statistical calculation found a rise of 10-30\% in the
expansion rate relative to the FRW value around a time of
10 billion years \cite{Rasanen:2008a, peakrevs}, which agrees
with the observations within an order of magnitude.
The calculation involved several approximations,
and a more careful treatment is needed.
In particular, the difference between relativistic
and Newtonian cosmology should be better understood to
isolate the relevant relativistic degrees of freedom,
related to spatial curvature and the electric and magnetic
components of the Weyl tensor.

\acknowledgments

I thank Thomas Buchert and Antonio Riotto for
comments on the manuscript. \\

\end{document}